# MOLECULAR DYNAMICS SIMULATION OF THE SOLIDIFICATION OF AlCoCuFeNi HIGH–ENTROPY ALLOY NANOWIRE

O.I. Kushnerov

*Oles Honchar Dnipro National University, Dnipro, Ukraine*
*e-mail: kushnrv@gmail.com*

**Molecular dynamics simulation of the solidification behavior of AlCoCuFeNi nanowire was carried out basing on the embedded atom potential with different cooling rates ($1\cdot10^{11}$, $1\cdot10^{12}$, and $1\cdot10^{13}$ K/s). To simulate an infinite nanowire, a periodical boundary condition along the nanowire axis direction was applied. The crystallization of the nanowire was characterized by studying the temperature dependence of the potential energy. The adaptive common neighbor analysis (CNA) was performed and the radial distribution function (RDF) was calculated to determine the structure and lattice parameters of phases of the AlCoCuFeNi nanowire. It has been shown that the final structure of investigated nanoparticle changes from amorphous to crystalline with decreasing of the rate of cooling.**

## 1. Introduction

Recently, high-entropy alloys (HEA) have attracted increasing attention [1]. Such alloys usually contain from 5 to 13 major elements in equiatomic or near equiatomic concentrations (5 to 35%). Because of the high entropy, usually simple substitutional solid solutions with BCC or FCC crystal lattices are formed during the solidification of multicomponent alloys. HEAs are characterized by unique structures and a number of excellent characteristics, such as hardness, wear-resistance, resistance to oxidation, corrosion, and ionizing radiation, high thermal stability and biocompatibility [1-3]. So, the HEAs are considered as promising materials for application in electronics, atomic power engineering, transportation equipment, space rocket hardware, medicine, etc.

Metallic nanowires are important for both technological applications as molecular electronic devices and fundamental low-dimensional physics. In order to understand the structures of metallic nanowires, it is interesting to investigate their structural evolutions during solidification, however, it is difficult to study the microstructures of nanowires directly or to observe the local structures because of the short length and time scales involved. Computer simulation provides a physical insight in understanding various phenomena on the atomic scale and enables one to predict some properties of nanomaterials. It has been shown that molecular dynamics simulations based on classical mechanics were able to reproduce the experimental results of deformation mechanisms and atomic motions in metals with sufficient accuracy and to provide an atomic description of the crystallization and glass-forming processes during rapid solidification of alloys. Also, molecular dynamics is one of the most powerful atomic simulation methods that is used to investigate the properties and structures of nanomaterials.

The present study is concerned with classical molecular dynamics (MD) simulation for studying the solidification of AlCoCuFeNi HEA nanowire.

## 2. Computational methods

In this research the embedded-atom method (EAM) [4] was used to describe the interaction among Al–Co–Cu–Ni–Fe. EAM potentials involve empirical formulas that simulate material characteristics such as the heat of solution, lattice constants, and surface energies. Such a potential is non-pairwise in the sense that it is based on concepts from density functional theory, which stipulate in general that the energy of a solid is a unique function of the electron density.

The EAM potential is expressed as a multi-body potential energy function in the following form [4]

$$E = \sum_i f_i(\bar{\rho}_i) + \frac{1}{2}\sum_i \sum_{j \neq i} \phi_{ij}(r_{ij}) \quad (1)$$

where $E$ is the total energy of the atomic system; it comprises the sum of the embedding energy (first term of Eq.(1)) and the short-range pair potential energy (the second term). Here, $f_i(\bar{\rho}_i)$ is the embedding energy function, $\phi_{ij}(r_{ij})$ is a pair interaction potential energy as a function of the distance $r_{ij}$ between atoms $i$ and $j$ that have chemical sorts $a$ and $b$, $\bar{\rho}_i$ is the local host electron density. In the present research, the EAM potential developed by Zhou et al. [5] was used.

Classical MD simulation studies of high entropy AlCoCuFeNi nanowire were performed using Large Scale Atomic/Molecular Massively Parallel Simulator (LAMMPS, Sandia National Laboratory, USA) [6]. Visualization of snapshots and the coordination analysis are done using open visualization tool software OVITO [7]. Simulations are performed under the *NVT* ensemble (constant number of atoms, the volume of the simulation cell and temperature), but the sample volume was not constant due to the fact that the sample was with a free surface (diameter of the nanowire was smaller than the simulation cell). Integration was done by the velocity – Verlet algorithm with a time step of 0.002 ps. The Nosé–Hoover thermostat was used in the MD simulation to perform temperature control. The simulated nanowire was composed of 50000 atoms (Al, Co, Cu, Fe, Ni in the equiatomic ratio), a diameter of the nanowire was ~7.1 nm and the length of the cell was 18.4 nm. To simulate an infinite wire, periodical boundary condition along the nanowire axis direction was applied. The system was heated up to 2300 K and subsequently annealed at this temperature for 200 ps to bring the system to an equilibrium state. Then, the nanowire was quenched from a molten state (2000 K) at a rate of $1\cdot10^{11}$ K/s (low), $1\cdot10^{12}$ K/s (medium), and $1\cdot10^{13}$ K/s (high) to 300 K. After quenching the radial distribution function (RDF) was calculated for determining the structure of nanowire. Also, the adaptive common neighbor analysis (CNA) was performed to identify the microstructure of the samples (on the basis of an algorithm determining the symmetry of the nearest-neighbor atom).

## 3. Results and discussion

The potential energy variations of AlCoCuFeNi nanowire at different cooling rates are shown in Fig.1. The transition from solid to the liquid phase could be identified by a jump in the energy curve (Fig. 1) [8]. Apparently, different cooling rates lead to different features. At cooling rates $1\cdot10^{12}$ K/s and $1\cdot10^{13}$ K/s, the potential energy decreases almost linearly with decreasing of the temperature, which indicates nonexistence of crystal in the whole cooling process and the final structure of the nanowire is expected to be amorphous. However, there is a drop in potential energy at the cooling rate of $1\cdot10^{11}$K/s. The reasons for this phenomenon may be explained by atomic diffusion. At high cooling rates ($1\cdot10^{12}$ K/s, $1\cdot10^{13}$ K/s), the atoms have not enough time to rearrange their position in the cooling process, therefore, the potential energy curve varies almost linearly with the temperature, and the final structure is amorphous. At low cooling rate ($1\cdot10^{11}$ K/s), the atoms have enough time to rearrange the positions, and then the crystallization happens in the range of 830–800 K. Comparing these values with the results obtained in [9] for a nanoparticle of the same material (885–855 K), a decrease in the crystallization temperature can be distinguished. This is obviously due to the smaller diameter of thenanowire in comparison with the nanoparticle. It should be noted that at a medium rate ($1\cdot10^{12}$ K/s), the atoms have some time to rearrange their position to obtain a more compact structure with lower values of potential energy (Fig.1). Thus, it can be noted that the final potential energies at 300 K tend to decrease along with the decrease of the cooling rates, which implies that there were different phase transformation routes along with the changes in the temperatures.

The RDF curves for crystallized at a rate of $1\cdot10^{11}$ K/s AlCoCuFeNi nanowire (at the temperature of 300 K) are given in Fig. 2. By the CNA analysis, it was established, that the simulated nanoparticle contains the FCC phase (56.6 %), HCP phase (26.8 %) and the unrecognized (Other) phase (16.5%), which, in accordance with RDF, has an amorphous structure. As for the BCC phase, it is practically absent (<0.2%) in the crystallized nanowire at a temperature of 300 K (in contrast to the results obtained for the nanoparticle [9]). The estimated from the RDF curves [10] lattice parameter of the FCC phase for the present MD simulations is 0.367 nm and for the HCP phase $a$=0.258 nm and $c$=0.447 nm. These values coincide with the results for the AlCoCuFeNi nanoparticle and the experimental data [9].

It should be noted that at the initial stages of the crystallization process and immediately after its completion, a rather large amount of the BCC phase is present in the nanowire (up to 16%). However, during the cooling of the crystallized sample to the temperature of 300 K, its structure is rearranged (Fig. 3a, b, amorphous phase is deleted).

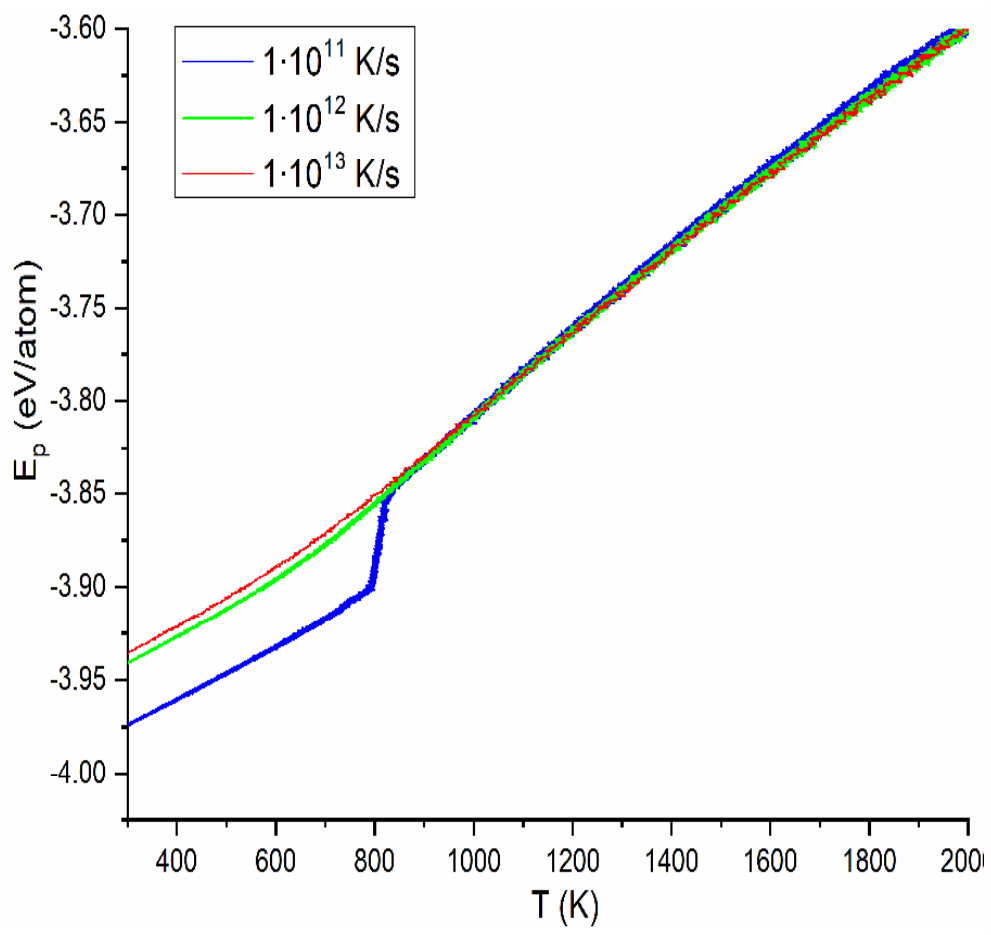

**Fig. 1. The temperature dependencies of the potential energy per atom of AlCoCuFeNi nanowire.**

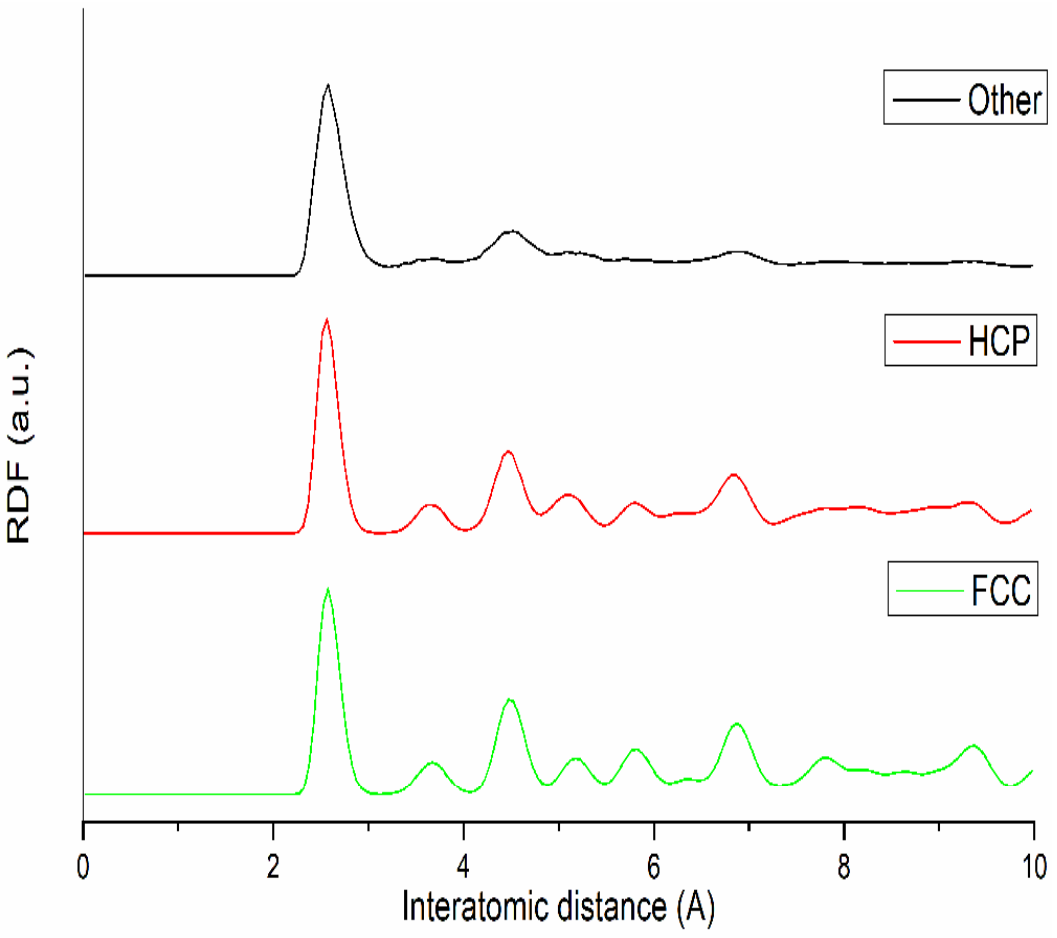

**Fig. 2. Total RDFs for AlCoCuFeNi nanowire crystallized at a rate of $1\cdot10^{11}$ K/s, individual curves correspond to different phases.**

The features of the distribution of the HCP phase atoms (Fig. 3c) allows us to conclude that this phase consists mainly of intrinsic (two adjacent HCP layers) and extrinsic (two HCP layers with an FCC layer between them) stacking faults in the lattice of the FCC phase [11]. The twin boundaries between the FCC twins (a single layer composed of HCP atoms) are also present [12]. As for the amorphous phase, its emergence can be explained by the high cooling rate at which some of the atoms have not enough time to rearrange and create a crystalline phase. Also in Fig. 3 distortions of the cylindrical shape of the nanowire caused by the growth of the crystalline phase are clearly visible. It should be noted that the similar arrangement of the twin boundaries and stacking faults, which form the slip system, takes place in the sample under uniaxial tensile loading (deformation twinning) [11].

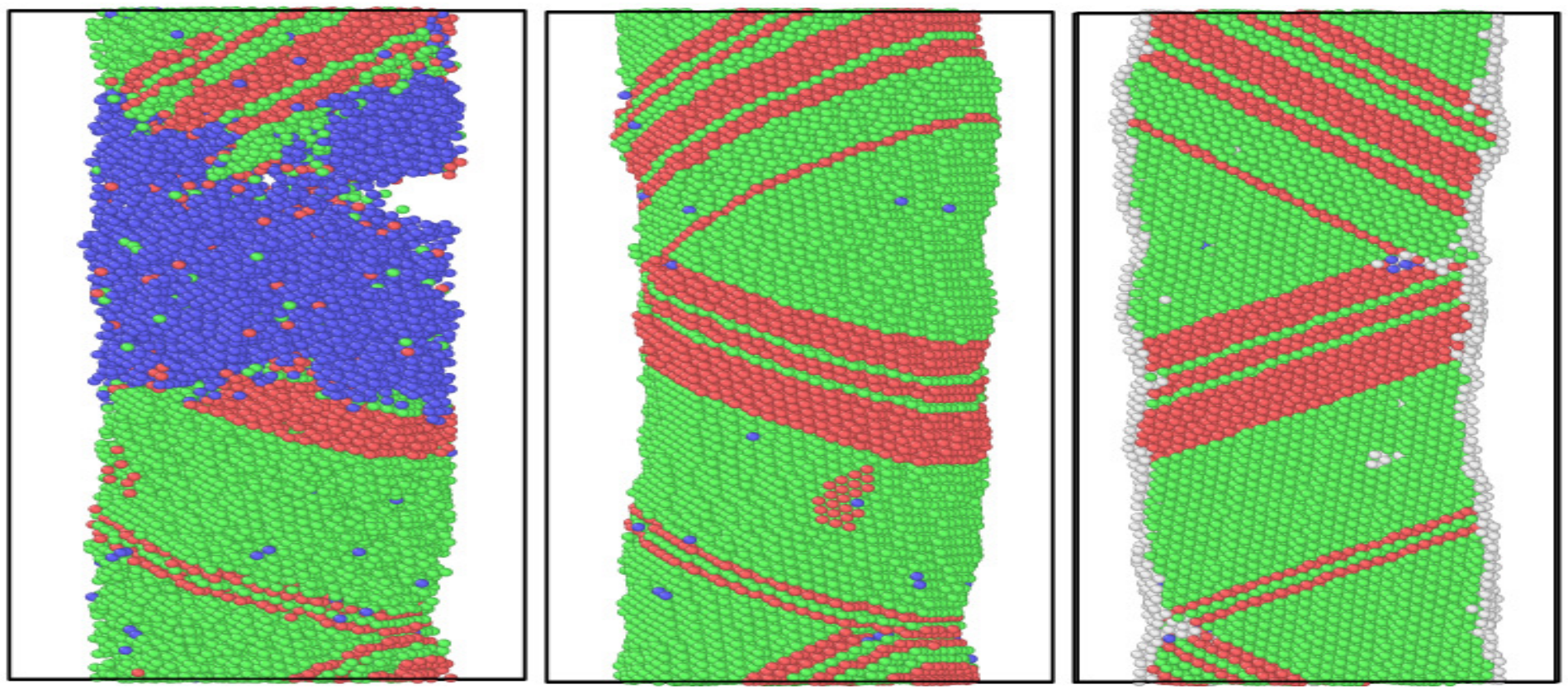

**Fig. 3. The microstructure of AlCoCuFeNi nanowire: a – at the temperature of 790 K, b – at the temperature of 300 K, c – longitudinal cross-section at the temperature of 300 K ( ● – FCC, ● – BCC, ● – HCP, ● – amorphous phase).**

## 4. Conclusions

The solidification processes of AlCoCuFeNi nanowire are studied by molecular dynamics simulation at three different cooling rates. It is found that the final structure of the investigated nanowire changes from amorphous to crystalline with decreasing the rate of cooling processes. From the simulation results, it is shown that the classical molecular dynamics simulation with the use of embedded atom potential allows us to describe structure and lattice parameters of AlCoCuFeNi nanowire in good agreement with experimental data. It is shown that the phase composition and crystallization temperature of a nanowire differs from those for an AlCoCuFeNi HEA nanoparticle.